\documentclass[12pt,preprint]{aastex631}

\usepackage{amsmath}
\usepackage{graphicx}
\usepackage{soul}

\newcommand{\bu}{{\bf u}}

\newcommand{\be}{{\bf e}}
\newcommand{\bk}{{\bf k}}
\newcommand{\bB}{{\bf B}}
\newcommand{\bb}{{\bf b}}
\renewcommand{\Pr}{\mathrm{Pr}}
\newcommand{\ad}{\mathrm{ad}}

\begin{document}

\author{A. Sanghi$^1$, A. E. Fraser$^1$, E. W. Tian$^1$, and P. Garaud}
\affil{
Department of Applied Mathematics, Baskin School of Engineering, \\
University of California, Santa Cruz, CA 95064\\
}
\title{Magnetized oscillatory double-diffusive convection}

\shorttitle{Magnetized ODDC}
\shortauthors{Sanghi et al.}

\begin{abstract}
    We study the properties of oscillatory double-diffusive convection (ODDC) in the presence of a uniform vertical background magnetic field. ODDC takes place in stellar regions that are unstable according to the Schwarzschild criterion and stable according to the Ledoux criterion (sometimes called semiconvective regions), which are often predicted to reside just outside the core of intermediate-mass main sequence stars. Previous hydrodynamic studies of ODDC have shown that the basic instability saturates into a state of weak wave-like convection, but that a secondary instability can sometimes transform it into a state of layered convection, where layers then rapidly merge and grow until the entire region is fully convective.  
    We find that magnetized ODDC has very similar properties overall, with some important quantitative differences. A linear stability analysis reveals that the fastest-growing modes are unaffected by the field, but that other modes are. Numerically, the magnetic field is seen to influence the saturation of the basic instability, overall reducing the turbulent fluxes of temperature and composition. This in turn affects layer formation, usually delaying it, and occasionally suppressing it entirely for sufficiently strong fields. Further work will be needed, however, to determine the field strength above which layer formation is actually suppressed in stars. Potential observational implications are briefly discussed.    
\end{abstract}

\section{Introduction}
\label{sec:intro}

 Oscillatory double-diffusive convection (ODDC hereafter) takes place in regions of stars or planets that have a destabilizing temperature stratification and a stabilizing composition stratification. These would be identified as convectively unstable according to the Schwarzschild criterion, but convectively  stable according to the Ledoux criterion\footnote{They are also often referred to as semiconvective regions \citep{Spiegel1969,Langer1983}, but since the terminology is contentious, we hereafter simply use ODDC, which refers to the fluid dynamical instability that takes place in regions with this type of stratification.}. In stellar evolution calculations that use the Ledoux-criterion for convection, regions with this type of stratification most commonly appear just above the convective core of intermediate-mass and high-mass stars. ODDC can in that case enhance the transport of chemical species between the envelope and the core, which provides additional fuel to the nuclear reactions and prolongs the Main Sequence phase \citep[see, e.g.][and references therein]{MooreGaraud2016}. 
 
  In its basic form, ODDC is a small-scale, wave-like  type of thermal convection. It was identified to be a double-diffusive instability by \citet{Kato1966}, and was hence named ODDC by \citet{Spiegel1969}. A key question is how much mixing ODDC can cause. The answer turns out to be quite complicated, because it depends on whether the instability remains in its basic small-scale form, or whether it  transforms into a state of large-scale layered convection, where mixing is much more efficient. Layered convection had been postulated to exist in giant planets and stars by \citet{Stevenson1985} and \citet{Spruit1992} based on its presence in geophysical analogs of ODDC \citep[e.g. in the polar oceans and volcanic lakes on Earth, see the review by][]{radko2013double}. But it was first clearly identified in direct numerical simulations (DNS hereafter) of ODDC at  astrophysically relevant input parameters only recently by \citet{rosenblumal2011}. 

In that work, \citet{rosenblumal2011} demonstrated that layering (i.e. transition to layered convection) can spontaneously occur in ODDC, and that it is caused by a mean-field instability first discovered by \citet{radko2003mechanism} in a related oceanographic context. Mean-field instabilities are defined mathematically as instabilities of the Reynolds-averaged equations, rather than of the original Navier-Stokes equations. Physically speaking, they can be viewed as secondary instabilities that only develop after a particular primary instability has saturated into a state of homogeneous turbulence. In addition, mean-field instabilities have characteristic lengthscales that are much larger than the typical turbulent eddy scale. Understanding their development requires knowledge (or modeling) of relevant turbulent fluxes such as the heat and composition fluxes (and others as needed) as functions of the large-scale properties of the fluid such as the background temperature gradient, the background composition gradient, and the mean local shear for instance. The specific mean-field instability discovered by \citet{radko2003mechanism} is called the $\gamma$-instability, and is triggered by a positive feedback loop between the heat and composition fluxes (whose ratio is traditionally called $\gamma$), and the background stratification \citep[see reviews by, e.g.][]{radko2013double,Garaud2018}. 

In subsequent investigations, \citet{Mirouh2012} ran and analyzed a wide range of DNS, to model the turbulent heat and composition fluxes associated with the basic weak form of ODDC. With that information, they were able to construct a quantitative model for the mean-field $\gamma$-instability that can be used to predict when layering is expected. A salient result of their analysis is that layering is always expected in ODDC-unstable regions located at the edge of the convective cores in intermediate-mass stars, suggesting that mixing in these regions is very efficient indeed. 

With this in mind, \citet{Woodal13} used DNS of layered convection to quantify transport in that regime. They found that both heat and composition fluxes increase with the mean layer height in a way that is strongly reminiscent of (but not entirely identical to) Rayleigh-B\'enard convection (i.e. thermal convection between plane parallel plates). But they also found that the layers have a propensity to merge with one another, in such a way that the ODDC-unstable region always eventually evolves into one that is fully convective. This suggests that the ODDC-unstable region is only expected to exist for a relatively short period of the star's life, and is rapidly replaced by a standard convection zone. 

Using their results, \citet{MooreGaraud2016}  modeled the evolution of ODDC-unstable regions in intermediate mass stars using MESA \citep{Paxtonal2011}, and found that mixing is so efficient that these regions can be approximated as being fully convective instead for simplicity. In other words, ignoring ODDC altogether and using the Schwarzschild criterion to identify convectively mixed regions satisfactorily predicts the star's evolution. A similar conclusion was recently reached by \citet{Andersal22}, albeit for different reasons. 

However, it is important to note that almost all results on ODDC so far have been obtained in the hydrodynamic limit \citep[although see ][for a related magnetized instability]{HughesBrummell21}, but core convection in rotating stars can drive a substantial dynamo \citep{Augustsonal16}. The magnetic field thus generated can diffuse out of the core and would likely influence the initial development of the ODDC instability, and/or its subsequent transition into layered convection. 

In this work, we are therefore interested in quantifying the effects of magnetic fields on ODDC. For simplicity, we begin by looking at a very simple model setup in which an ODDC-unstable region is threaded by a uniform, vertical magnetic field. Section \ref{sec:model} describes the governing equations for the problem. Section \ref{sec:linear} presents a linear stability analysis of ODDC in the presence of this field, demonstrating that the fastest-growing modes of instability are unaffected. Section \ref{sec:num} then presents nonlinear DNS of the system, where in this case the magnetic field is seen to slow down, and sometimes even suppress, the transition to layered convection. We interpret our findings in the light of Radko's $\gamma$-instability theory in Section \ref{sec:gamma}, and conclude in Section \ref{sec:ccl} with a discussion of the impacts of our findings on stellar modeling.





\section{Mathematical Model and Governing Equations}
\label{sec:model}

We consider a Cartesian domain in a small stellar region with dimensions  $(L_x,L_y,L_z)$ where gravity defines the vertical direction ${\bf e}_z$. We assume that the vertical extent of the domain is smaller than a pressure scale height, which allows us to use the Boussinesq approximation \citep{SpiegelVeronis1960}. Within this approximation, we assume that the temperature $T$ and composition $C$ can be decomposed as a linear background plus perturbations, namely
\begin{eqnarray}
T(x,y,z,t) = T_m + z \frac{dT_0}{dz} + \tilde{T}(x,y,z,t),  \\
C(x,y,z,t) = C_m + z \frac{dC_0}{dz} + \tilde{C}(x,y,z,t),
\end{eqnarray}
where $x$, $y$, and $z$ are the spatial coordinates and $t$ is time, $T_m$ and $C_m$ are constants that represent the mean values of temperature and composition in the region considered, and $dT_0/dz$ and $dC_0/dz$ are negative constant gradients of temperature and composition, with $L_z |dT_0/dz| \ll T_m$ and $L_z |dC_0/dz| \ll C_m$. Consistent with the use of the Boussinesq approximation, we assume that the density perturbations ${\tilde\rho}$ satisfy a linear equation of state:
\begin{eqnarray}
&& \frac{\tilde\rho}{\rho_m}  = -\alpha\tilde{T} + \beta\tilde{C}  
\end{eqnarray}
where $\rho_m$ is the mean density of the domain, and $\alpha= -\frac{1}{\rho_m}\frac{\partial \rho}{\partial T}$ and $\beta = \frac{1}{\rho_m}\frac{\partial \rho}{\partial C}$ are constants of thermal expansion and compositional contraction respectively. 

 With these assumptions, and neglecting magnetic buoyancy effects, the governing equations are:
\begin{eqnarray}
&& \nabla \cdot \bu = 0, \nonumber\\
&& \nabla \cdot \bB = 0, \nonumber\\
&& \rho_m \left( \frac{\partial \bu}{\partial t} + \bu \cdot \nabla \bu \right) =  - \nabla \tilde{p} - \tilde{\rho} g{\bf e}_z + \frac{1}{\mu_0} ( \nabla \times \bB)\times \bB  + \rho_m \nu \nabla^2 \bu, \nonumber\\
&&  \frac{\partial \tilde{T}}{\partial t} + \bu \cdot \nabla \tilde{T} + u_z \left(\frac{dT_0}{dz} - \frac{dT_{\ad}}{dz}\right)=  \kappa_T \nabla^2 \tilde{T}, \nonumber \\
&& \frac{\partial \tilde{C}}{\partial t} + \bu \cdot \nabla \tilde{C} + u_z\frac{dC_0}{dz}=  \kappa_C \nabla^2 \tilde{C}, \nonumber \\
&& \frac{\partial \bB}{\partial t} = \nabla \times (\bu \times \bB) +\eta \nabla^2\bB,
\end{eqnarray}
where $\bu=(u_x,u_y,u_z)$ is the velocity of the fluid, $\bB=(B_x,B_y,B_z)$ is the magnetic field,  $g$ is gravity, $\mu_0$ is the vaccuum permeability, and $\tilde{p}$ is the pressure perturbation away from hydrostatic equilibrium. 
The diffusion coefficients, namely the kinematic viscosity $\nu$, the thermal diffusivity $\kappa_T$, the compositional diffusivity $\kappa_C$ and the magnetic diffusivity $\eta$ are all assumed to be constant. Finally, $dT_{\ad}/dz = -g/c_p$ is the vertical adiabatic temperature gradient, where $c_p$ is the specific heat at constant pressure of the fluid. To be considered Schwarzschild-unstable and Ledoux-stable the region satisfies
\begin{equation}
    \frac{dT_0}{dz} - \frac{dT_{\ad}}{dz} < 0, \mbox{  and   } \frac{dT_0}{dz} - \frac{dT_{\ad}}{dz} - \frac{dC_0}{dz } > 0. 
\end{equation}

These equations must be complemented by boundary conditions. Two options are possible: to run ``global" simulations in bounded domains with boundary conditions enforced on the dynamical fields \citep[e.g.][]{Andersal22}, or to run ``local" simulations where background gradients are fixed and periodic boundary conditions are enforced for the perturbed fields $\tilde{T}$, $\tilde{C}$, $\bu$, and $\bB$ \citep[e.g.][]{rosenblumal2011,Mirouh2012,Woodal13}. The regions in stellar interiors where ODDC might occur are far from boundaries; thus, to minimize their effect on our results, we perform local simulations with periodic boundary conditions. This numerical setup is meaningful and valid as long as two conditions are met. First, the domain size must be larger than the dominant eddy length scales in the saturated state to ensure that the boundaries are not unphysically influencing the turbulence. Second, we note that the periodic model implicitly maintains large-scale temperature and composition jumps across the height of the domain, letting the turbulence develop accordingly. This is only a good approximation to the true physical problem if the timescale over which the turbulence would in reality modify (and reduce) both jumps is long compared with the timescale for the development and nonlinear saturation of the instability. This was shown to be the case for the related fingering instability by \citet{Zemskova2014}, and we assume that it holds here as well for ODDC. This assumption is very likely justified in the case where ODDC remains in its weak, small-scale, wave-like form, but could be invalidated in the layered case (where convective fluxes transport both heat and composition very rapidly). Studying the transition to layered convection in a different model setup is the subject of a forthcoming paper.

We non-dimensionalize the system using the expected width of diffusive structures \textit{d} given by \citep{stern1960sfa} 
\begin{eqnarray}
&&[l]=d = \left(\frac{\kappa_T \nu}{\alpha g\left|\frac{dT_0}{dz} - \frac{dT_{\ad}}{dz}\right|}\right)^{1/4} = \left(\frac{\kappa_T \nu}{N_T^2}\right)^{1/4}, 
\end{eqnarray}
where $N_T$ is the local buoyancy frequency associated only with the temperature field. The corresponding choices for the other units are: 
\begin{eqnarray}
&&[t]= \frac{d^2}{\kappa_T}, \quad \quad [u]=\frac{\kappa_T}{d}, \nonumber\\
&&[T]= d\left|\frac{dT_0}{dz} - \frac{dT_{\ad}}{dz}\right|, \quad \quad [C]= \frac{\alpha}{\beta} d\left|\frac{dT_0}{dz} - \frac{dT_{\ad}}{dz}\right|.
\end{eqnarray}
Finally, we assume the existence of a constant background field of amplitude $B_0$, so 
\begin{equation}
    {\bf B} = B_0 {\bf e}_z + {\bf b}, 
\end{equation}
and we use this amplitude as the unit for the magnetic field strength. With this, the non-dimensional system of governing equations is:
\begin{eqnarray}
&& \nabla \cdot \hat{\bu} = 0, \quad \nabla \cdot\hat \bB = 0 ,  \label{eq:incomp}\\
&&\frac{\partial \hat{\bu}}{\partial t} + \hat{\bu} \cdot \nabla \hat{\bu} =  -\nabla \hat{p} + \Pr \nabla^2 \hat{\bu} + H_B (\nabla \times \hat{\bB})\times \hat{\bB}+ \Pr\left(\hat{T}-\hat{C}\right) \be_z, \\
&& \frac{\partial \hat{T}}{\partial t} + \hat{\bu} \cdot \nabla \hat{T} - \hat{u}_z  = \nabla^2 \hat{T},  \label{eq:tempeq}\\
&& \frac{\partial \hat{C}}{\partial t} + \hat{\bu} \cdot \nabla\hat{C} -R_0^{-1}\hat{u}_z  =  \tau \nabla^2 \hat{C} ,  \label{eq:Ceq}\\
&& \frac{\partial \hat{\bB}}{\partial t} = \nabla \times (\hat{\bu} \times \hat{\bB}) +D_B \nabla^2\hat{\bB}, \label{eq:induc}
\end{eqnarray}
where hatted quantities, as well as time and space, are now non-dimensional. Several parameters appear, namely the so-called inverse density ratio which measures the stratification of the region,
\begin{equation}
    R_0^{-1} = \frac{\beta| dC_0/dz|}{\alpha|dT_0/dz - dT_{\ad}/dz|}, 
\label{eq:R0invdef}
\end{equation}
three diffusivity ratios which only depend on the properties of the fluid, 
\begin{equation}
    \Pr = \frac{\nu}{\kappa_T}, \quad \tau = \frac{\kappa_C}{\kappa_T}, \mbox{ and }  D_B = \frac{\eta}{\kappa_T}, 
\end{equation}
and a parameter controlling the magnetic field strength,
\begin{equation}
H_B = \frac{B_0^2 d^2}{\rho_m \mu_0
\kappa_T^2}, 
\end{equation}
which is the square of the ratio of the Alfv\'en velocity to the anticipated ODDC velocity $\kappa_T / d$. 

Note that in the hydrodynamic case, a region is linearly unstable to ODDC provided 
\begin{equation}
    1 < R_0^{-1} < \frac{\Pr + 1}{\Pr + \tau} \equiv R_c^{-1},
\end{equation}
linearly stable if $R_0^{-1} > R_c^{-1}$, and unstable to thermosolutal convection if $R_0^{-1} < 1$ \citep{baines1969}. In stellar interiors, the diffusivity ratios are always quite small \citep[cf.][]{Garaudal2015}, so $R_c^{-1}$ is very large. Also note that an estimate for $H_B$ near the core of intermediate-mass stars is 
\begin{equation}
    H_B = \left(\frac{B_0}{10^4 \mbox{G}}\right)^2 \left( \frac{d}{10^3 \mbox{cm}}  \right)^2  \left(\frac{\rho_m}{100 \mbox{g/cm}^3} \right)^{-1} 
    \left(\frac{\kappa_T}{10^6 \mbox{cm}^2/\mbox{s}}\right)^{-2},
\end{equation}
showing that a magnetic field with amplitude of about $10^4$G can in principle have a substantial impact on ODDC. 

The system of equations (\ref{eq:incomp})-(\ref{eq:induc}) is identical to the one used by \citet{HarringtonGaraud2019} to study magnetized fingering convection, a related double-diffusive instability, except for the negative signs in front of the terms $\hat u_z$ and $R_0^{-1} \hat u_z$ in the temperature and composition equations respectively (see their equations 14-18). 

 \section{Linear stability analysis} 
 \label{sec:linear}
 
We analyze the stability of the system (\ref{eq:incomp})-(\ref{eq:induc}) to infinitesimal perturbations following standard procedures. First, we 
explicitly write $\hat {\bf B} = {\bf e}_z + \hat {\bf b}$, where $\hat {\bf b}$ are magnetic field perturbations. The background state around which we linearize is a steady state with $\hat {\bf u} = \hat {\bf b} = \hat T = \hat C = \hat p = 0$. We assume the perturbations to be sufficiently small so that nonlinearities can be ignored. 
 Finally, we seek solutions of the form $\hat{q} = q'\exp(i\hat{k}_x x + i \hat{k}_y y +  i\hat{k}_z z+ \hat{\lambda} t)$ where $\hat{\lambda}$ is the growth rate of an unstable mode with wavenumber $\hat{\bk} = (\hat{k}_x, \hat{k}_y, \hat{k}_z)$. 
With these steps and assumed ansatz, the governing equations become: 
\begin{eqnarray}
&& \hat{\bk} \cdot \bu' = 0, \quad \quad  \hat{\bk} \cdot \bb' = 0, \\
&& (\hat{\lambda}+ \Pr \hat{k}^2 ) \bu' =  - i \hat{\bk} p' + i H_B(\hat{\bk} \times \bb') \times {\bf e}_{z} + \Pr(T' - C')  {\bf e}_{z},  \\
&& ( \hat{\lambda}+ \hat{k}^2) T'   =  u'_z ,  \\
&& ( \hat{\lambda}+\tau \hat{k}^2) C'  =  R_0^{-1} u'_z , \\
&& (\hat{\lambda} + D_B \hat{k}^2)\bb'= i\hat{\bf k} \times ( \bu' \times {\bf e}_z) ,
\end{eqnarray}
where $\hat k^2 = \hat k_x^2 + \hat k_y^2 + \hat k_z^2 = \hat k_h^2 + \hat k_z^2$ and $\hat k_h$ is a horizontal wavenumber. 
The linearized induction equation can be simplified as
\begin{equation}
\bb' = \frac{i\hat{k}_{z}\bu'}{\hat{\lambda} + D_B\hat{k}^2},
\label{eq:linearind}
\end{equation}
while the temperature and composition equations can be combined into
\begin{equation}
T'-C' = u'_z\left( \frac{1}{\hat{\lambda} + \hat{k}^2} - \frac{R_0^{-1} }{ \hat \lambda + \tau \hat{k}^2}\right) .
\label{eq:linearTC}
\end{equation}
Substituting (\ref{eq:linearind}) and (\ref{eq:linearTC}) into the linearized momentum equation, taking the dot product of the result with $\hat{\bf k}$, and using incompressibility, yields an expression for $p'$. After substituting it back into the $z$-component of the momentum equation, we finally obtain a quartic equation for $\hat{\lambda}$: 
\begin{eqnarray}
&&(\hat \lambda +\hat k^2) (\hat \lambda + \tau \hat k^2) \left(\hat \lambda + \Pr \hat k^2 + \frac{H_B\hat k_{z}^2}{\hat \lambda + D_B\hat k^2}\right)  = \Pr\left(  (\hat \lambda + \tau \hat k^2) - R_0^{-1} (\hat \lambda + \hat k^2)\right) \frac{\hat k_h ^2}{\hat k^2}.
\label{eq:quartic}
\end{eqnarray}

This quartic has several notable properties. We see that setting $H_B = 0$ recovers the standard cubic associated with hydrodynamic ODDC, see, e.g. \citet{rosenblumal2011}. In addition, the only difference with the quartic obtained by \citet{HarringtonGaraud2019} describing the linear stability of magnetized fingering convection is the sign of the  right-hand side of this equation. As such, many of their results hold here as well. For instance, the magnetic field has no effect on the stability of modes with $\hat k_{z} = 0$ (i.e. modes that are invariant along both gravity and the background magnetic field). This is expected, since magnetic fields only interact with flows perpendicular to the field lines. However, since the $\hat k_{z} = 0$ modes are also the most rapidly growing modes when $H_B = 0$, we conclude that the presence of a vertical magnetic field does not change the stability of the most rapidly growing modes.\footnote{While $\hat{k}_z = 0$ modes are somewhat unphysical in that they are only permissible in local domains with periodic boundary conditions, they are reasonable approximations of the fastest-growing modes that would actually be realized in stellar interiors, where large-scale effects such as differential rotation or radial structure variation set an effective minimum wavenumber so small that it is effectively zero for all practical purposes.}

Writing (\ref{eq:quartic}) in the form $\hat \lambda^4 + \alpha_3\hat \lambda^3 + \alpha_2\hat \lambda^2 + \alpha_1\hat \lambda + \alpha_0 = 0$ gives the following expressions for the coefficients $\alpha_{j}$: 
\begin{eqnarray}
&& \alpha_3 = \hat k^2(\Pr + D_B + \tau + 1), \nonumber\\
&& \alpha_2 = \hat k^2_{z}H_B + \hat k^4(\Pr + D_B)(1+\tau)+  \hat k^4\Pr D_B + \hat k^4\tau + \Pr\frac{\hat k^2_x}{\hat k^2}(R_0^{-1}-1),\nonumber\\
&& \alpha_1 = \hat k^2(1+\tau)(\hat k^4D_B \Pr + \hat k^2_{z}H_B) + \hat k^6 (\Pr + D_B)\tau + \Pr \hat k^2_x (D_B(R_0^{-1}-1) + R_0^{-1}-\tau),\nonumber\\
&& \alpha_0 = \tau \hat k^4(\hat k^4D_B \Pr + \hat k^2_{z}H_B) + \Pr \hat k^2_x \hat k^2 D_B(R_0^{-1}-\tau),\nonumber
\end{eqnarray}
where we have set $\hat k_h = \hat k_x$ without loss of generality.

\begin{figure}
    \centering
    \includegraphics[width = \textwidth]{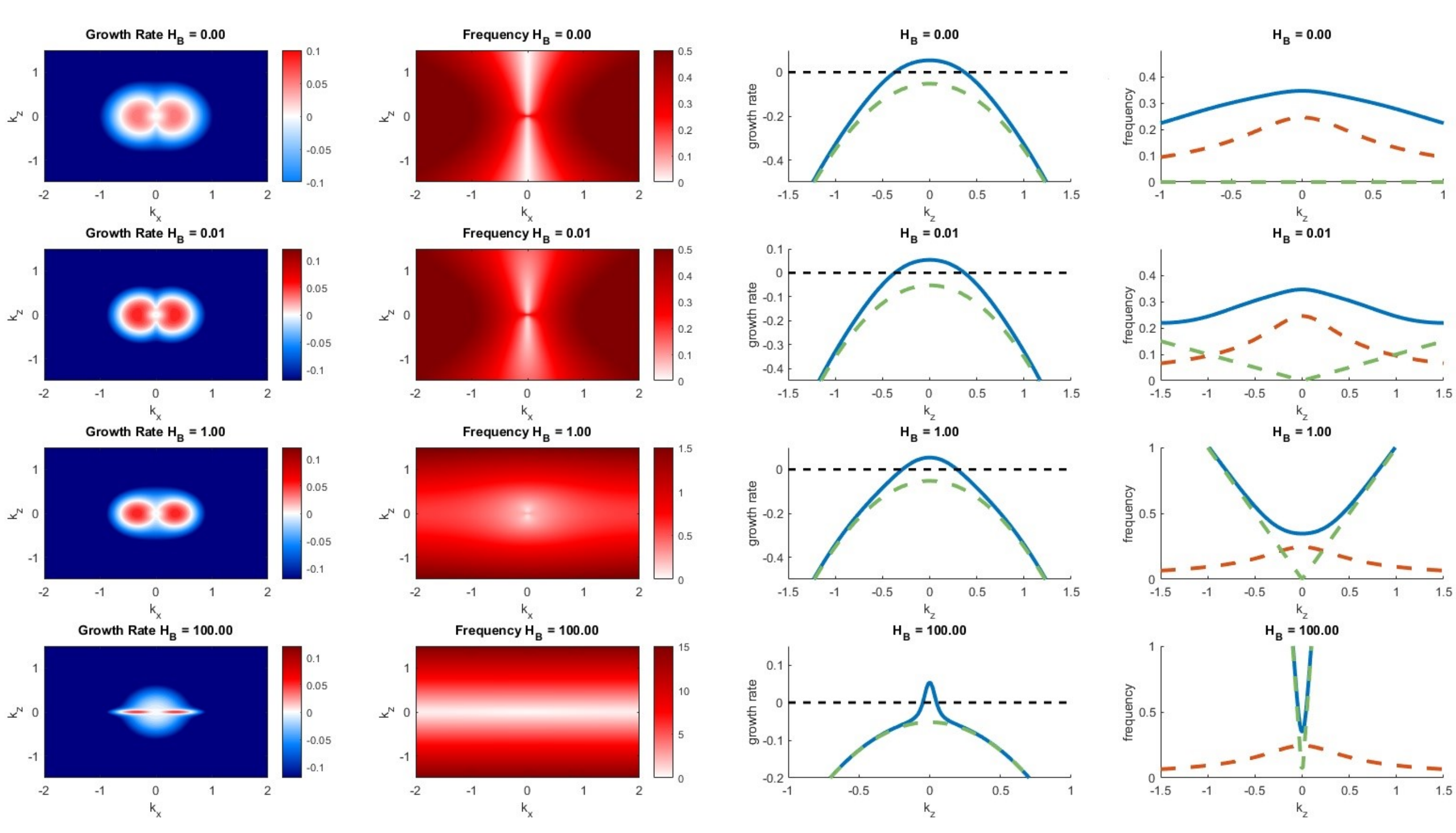}
    \caption{Properties of the growth rate $\hat \lambda$ of the ODDC instability for $\Pr$ = $\tau$ = $D_B$ = 0.3, $R_0^{-1} = 1.2$. $H_B$ increases from 0 to 100 from the top row to the bottom row. The first column shows $Re(\hat \lambda)$ as a function of $\hat{k}_z$ and $\hat{k}_x$. The second column similarly shows $|Im(\hat \lambda)|$. The third and fourth columns show $Re(\hat \lambda)$ and $|Im(\hat \lambda)|$, respectively, at fixed $\hat k_{x} = 0.417$ (the fastest growing mode at $\hat k_z = 0$). In the third column, $Re(\hat \lambda)$ (blue solid line) is compared with the diffusive solution $-c \hat k^2$ (green dashed line), see the main text for details. In the fourth column, $Im(\hat \lambda)$ (blue solid line) is compared with the oscillation frequencies of corresponding gravity waves $\omega_g$ (brown dashed curve) and corresponding Alfv\'en waves $\omega_A$ (green dashed curve).}
    \label{fig:linstab}
\end{figure}

We solve this quartic for $\hat \lambda$ for given input parameters ($R_0^{-1}$, $H_B$, $\Pr$, $\tau$ and $D_B$), and given values of the wavenumber $\hat {\bf k}$. Out of the four complex roots, we select the one with the maximum real part. From here on, we only ever discuss the properties of this particular root. We present the results of the linear stability analysis in Figure \ref{fig:linstab}. Each row corresponds to a different magnetic field strength, namely $H_{B} = 0, 0.01, 1, 100$, from top to bottom. From left to right in each column, we present (a) the real part of $\hat \lambda$ as a function of $\hat{k}_x$ and $\hat{k}_z$ (b) the imaginary part of $\hat \lambda$ as a function of $\hat{k}_x$ and $\hat{k}_z$ (selecting for simplicity the root with positive imaginary part), (c) the real part of $\hat \lambda$ as a function of $\hat{k}_z$, at $\hat{k}_x$ = 0.417, and (d) the imaginary part of $\hat \lambda$ as a function of $\hat{k}_z$, at $\hat{k}_x$ = 0.417. Note that the color scales on the first and second columns are quite different, and that $\hat k_x = 0.417$ is the fastest growing mode at $\hat{k}_z=0$ for the given parameters. In the third column, we compare $Re(\hat \lambda)$ to a purely diffusive solution of the form $-c(\hat{k}_x^2 + \hat{k}_z^2)$, where $c$ is the minimum of $\Pr, \tau$, and $D_B$. In the last column, we compare $|Im(\hat \lambda)|$ to the oscillation frequency of  a corresponding gravity wave, namely 
\begin{equation}
    \omega_g  = N \frac{|\hat k_x|}{\sqrt{\hat k_x^2 + \hat k_z^2}} = \sqrt{\Pr(R_0^{-1}-1)}\frac{|\hat k_x|}{\sqrt{\hat k_x^2 + \hat k_z^2}} ,
\end{equation}
and to the frequency of a corresponding Alfv\'en waves, namely 
\begin{equation}
    \omega_A = \sqrt{H_{B}} |\hat k_z |.
\end{equation} 

In general, and consistent with the discussion above, we see in the left-most column that the magnetic field partially or fully stabilizes every mode except those with $\hat k_z = 0$. More specifically, we see that at a given (non-zero) value of $\hat{k}_x$, increasing the magnetic field strength reduces the range in $\hat{k}_z$ of unstable modes, and reduces the growth rates of all modes except $\hat{k}_z =  0$. 

 In the second column, we see that as the value of $H_{B}$ increases, the oscillation frequency of the unstable modes becomes gradually influenced by the presence of the field. Indeed, in the first two rows of the last column, the field is zero or weak, and we find that the oscillation frequency of the $\hat{k}_z \ne 0$ modes are related to (but not equal to) the oscillation frequency of a pure gravity wave $\omega_g$. For larger values of $H_{B}$ (third and fourth row, where $H_B \ge 1$), the oscillation frequency of the modes approaches the corresponding Alfv\'en frequency $\omega_A$ instead, implying that the magnetic field now dominates the dynamics of the instability. 

A cursory examination of the decay rate of the decaying modes in the third column reveals that the latter has a parabolic dependence on $\hat k_z$ for large enough $\hat k_z$. This can be explained by noting that for large wavenumber,  the quartic (\ref{eq:quartic}) asymptotically tends to: 
\begin{equation}
\begin{split}
\hat \lambda^4 + \hat k^2(\Pr + D_B + \tau + 1)\hat \lambda^3 + \hat k^4[(\Pr + D_B)(1+\tau)+  \Pr D_B +\tau]\hat \lambda^2 + \\
\hat k^6[(\Pr + D_B)\tau + D_B \Pr(1 +\tau)]\hat \lambda + \hat k^8(D_B\Pr + \tau) = 0.
\end{split}
\end{equation}
This can be factored as $(\hat\lambda - \Pr \hat k^2)(\hat\lambda - \tau \hat k^2)(\hat\lambda - D_B 
\hat k^2)(\hat\lambda - \hat k^2) = 0$, which has roots $\hat\lambda = -c\hat k^2$, where $c$ can be 1, $\Pr, \tau$ or $D_B$. The slowest decaying mode is therefore obtained by taking $c = {\rm min}(1, Pr, \tau, D_B)$. The parabola $\hat\lambda = -c \hat k^2$ is shown in green in the third column, and we see that the actual solution of the quartic (\ref{eq:quartic}) (blue curve) tends to the diffusive solution (green curve) at large values of $\hat k_{z}$. In stars, we expect $c = \tau$, as we almost always have the ordering of parameters $\tau < \Pr < D_B < 1$. 


To summarize, we have found that a vertical magnetic field has no effect on the growth rate and nature of the fastest-growing, $z-$invariant mode of instability in this model setup. It does, however, affect both the growth rate and oscillation frequency of inclined modes, that presumably play a role in the saturation of the instability.  

\section{Numerical Simulations}
\label{sec:num}

To study the behavior of magnetized ODDC beyond the initial growth of the instability, and determine whether magnetic fields affect the processes by which convective layers form (see Section \ref{sec:intro}), we now turn to DNS. Equations (\ref{eq:incomp})-(\ref{eq:induc}) are evolved with time using the PADDIM code developed by \citet{HarringtonGaraud2019}. PADDIM itself is based on 
the original pseudo-spectral PADDI code  \citep{Traxler2011a,Stellmach2011} used in our previous non-magnetic studies \citep{rosenblumal2011,Mirouh2012,Woodal13}. 

We select a parameter regime where layers are known to form in the absence of magnetic fields, namely $\Pr = \tau = 0.3$, and $R_{0}^{-1} = 1.2$ \citep{Mirouh2012}, and run three simulations with increasing background magnetic field strength: $H_{B} = 0$ (hydrodynamic reference case), $H_B = 0.03$, and $H_B = 0.1$. The magnetic diffusion coefficient is fixed and equal to $D_B = 0.3$. Initial conditions in each simulation are: 
\begin{eqnarray}
&&\hat B_x = \hat B_y = 0, \hat B_z = 1 \mbox{ (for the magnetic simulations only)}, \nonumber\\
&& \hat u_x = \hat u_y = \hat u_z = 0, 
\end{eqnarray}
and both temperature and composition fields are initialized with small-amplitude white noise. In each case, the domain size used has size $100d \times 100d \times 300d$, and the resolution is $192 \times 192 \times 576$ equivalent grid points. 

\begin{figure}
    \centering
    \includegraphics[width=0.95\textwidth]{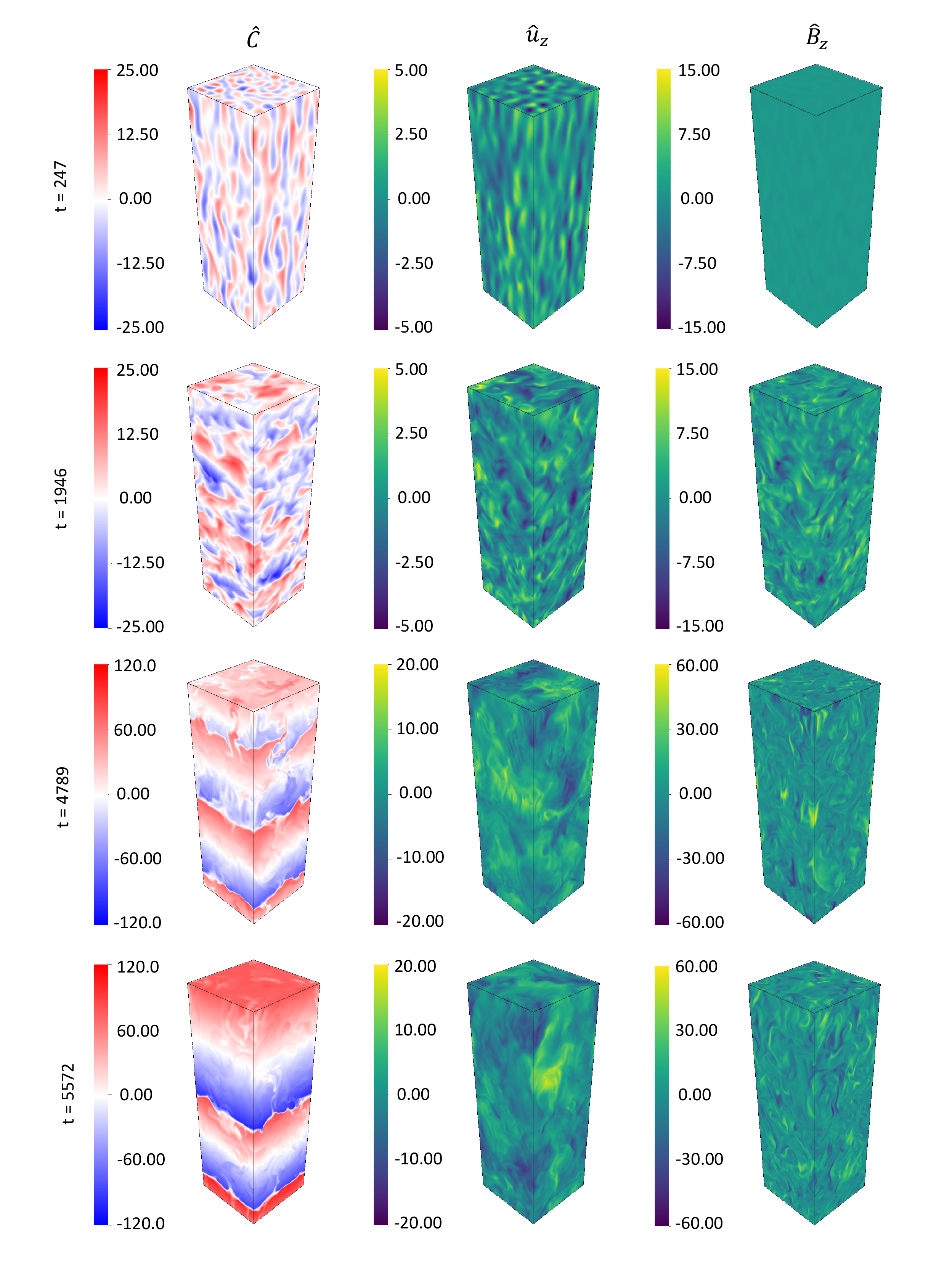}
\caption{Snapshots of the composition field $\hat C$, vertical Velocity field $\hat u_z$, and vertical magnetic field $\hat B_z$, in a simulation with parameters $\Pr = \tau = D_B = 0.3$, $R_{0}^{-1} = 1.2$, and $H_B = 0.03$. Each row corresponds to a different timestep, with time increasing (as marked in the figure) from top to bottom.
}
\label{fig:fluxes_layers}
\end{figure}

Typical snapshots from the simulation with $H_B = 0.03$ are shown in Figure \ref{fig:fluxes_layers}, illustrating the evolution of the system. Note that this would correspond to a weak-field case according to linear theory. The left column shows the composition field $\hat C$, the middle column shows the vertical velocity field $\hat u_z$, and the right column shows the vertical magnetic field $\hat B_z$. The top row shows the early evolution of the instability and the emergence of the oscillating vertically-invariant modes. These modes quickly saturate in the second row into a state that is dominated by weakly nonlinear waves, which is what is commonly called homogeneous oscillatory double-diffusive convection.    
Later in the simulation (third row) a stack of convective layers emerge, and then successively merge (fourth row), until a single layer remains (not shown). The amplitude of the flow velocity and of the magnetic field perturbations is much larger in the layered phase than in the homogeneous phase, consistent with the findings of \citet{rosenblumal2011} and \citet{Woodal13}. Layers are much more visible in the composition field than in the vertical velocity and magnetic field. This is because the density contrast across the interface is weak at the parameters selected, and is therefore not a strong barrier to vertical fluid motions. Instead, the interface is very turbulent and regularly pierced by strong upflows or downflows \citep[cf.][]{Woodal13}.

\begin{figure}
\includegraphics[width = \textwidth]{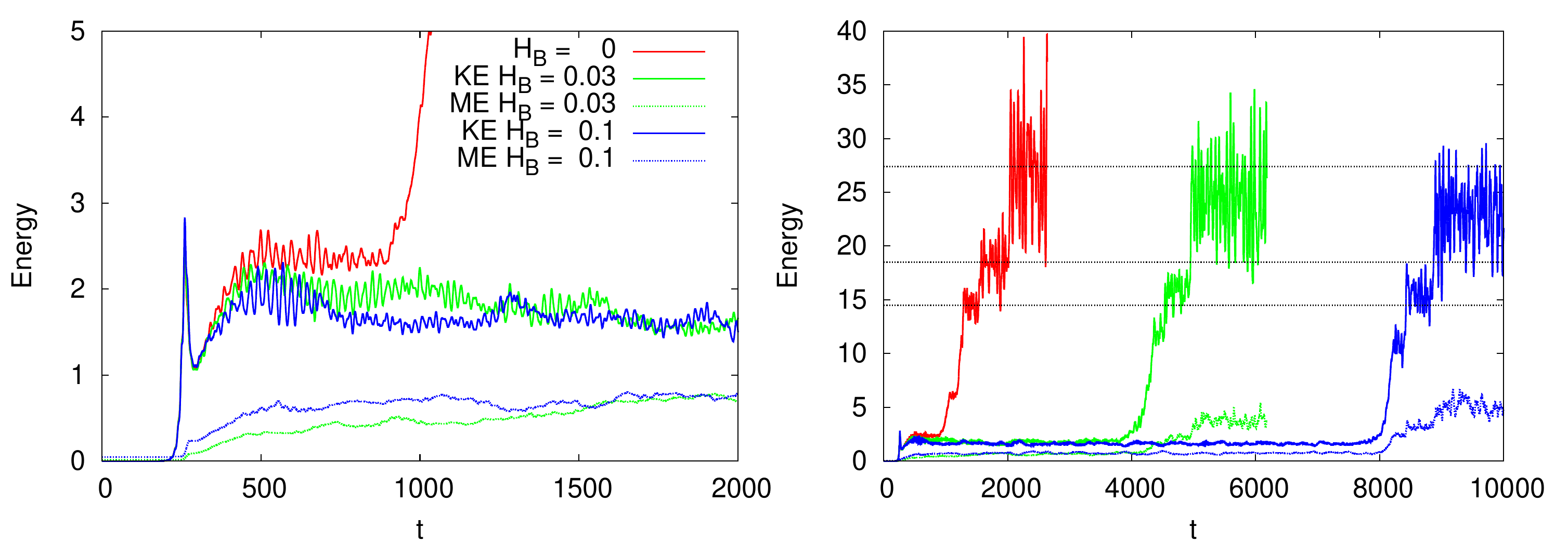}
\caption{Temporal evolution of the kinetic energy density (KE, solid line) and magnetic energy density (ME, dashed line) for three different values of the magnetic field strength: $H_B = 0$ (red line), $H_B = 0.03$ (green line) and $H_B= 0.1$ (blue line), at fixed $\Pr = \tau = D_B = 0.3, R_{0}^{-1} = 1.2$. The left panel shows the early-time behavior, while the right panel shows the entire evolution.
The horizontal black lines on the right panel show the kinetic energy density of the non-magnetic simulation measured, from bottom to top, in the four-layer phase, the three-layer phase, and the two-layer phase, to help compare it to that of the magnetized simulations.}
\label{fig:energies}
\end{figure}

A more quantitative way of looking at the evolution of the flow, which allows us to compare the outcome of simulations run at different magnetic field strengths, is presented in Figure \ref{fig:energies}. This figure shows the temporal evolution of the 
kinetic energy density $KE$ and magnetic energy density $ME$, defined as
\begin{equation}
    KE(t) = \frac{1}{2} \langle \hat u_x^2 + \hat u_y^2 + \hat u_z^2 \rangle, \quad ME(t) = \frac{H_B}{2} \langle \hat B_x^2 + \hat B_y^2 + \hat B_z^2 \rangle, 
 \end{equation}
where the angled brackets denote a volume average. The left-side panel focuses on the early-time behavior of the energies, while the right-side panel shows the entire time evolution. 

For $H_B = 0$, the simulation behaves as described by \citet{Mirouh2012}. The primary ODDC instability develops rapidly, as can be seen by the exponential growth of the kinetic energy, and saturates at a low amplitude around $t \simeq 250$. The kinetic energy grows again between $t \simeq 250$ and $t \simeq 500$, as a result of the excitation of larger-scale gravity waves that then saturate at a fairly constant amplitude between $t \simeq 500$ and $t \simeq 1000$. During that time, a stack of seven layers forms as the result of an underlying mean-field $\gamma$-instability (see next section for details). This is accompanied by a step-like increase in the kinetic energy (around $t \simeq 1000$), which corresponds to the onset of overturning convection in each layer. The seven layers very quickly merge into five, which then more slowly merge into four, three and then two layers around times $t \simeq 1200$, $t \simeq 1500$ and $t \simeq 2000$, respectively. Had the simulation been carried out for longer, we anticipate that the last two layers would also eventually merge. Each successive merger is accompanied by another strong step-wise increase in the kinetic energy. This is consistent with the findings of \citet{Woodal13}, who noted that the efficiency of layered convection increases significantly with the average layer height. 
 
We find that the presence of a vertical magnetic field with $H_B = 0.03$ and $0.1$, for the chosen parameters ($Pr = \tau = D_B = 0.3$, $R_0^{-1} = 1.2)$ has substantial quantitative effects on both the initial saturated state of the instability and on the development and evolution of  layered convection, despite being relatively weak from the perspective of linear theory. More specifically, 
we see that the magnetic field can reduce the kinetic energy in the flow prior to layer formation substantially (e.g. comparing energy levels in the phase prior to layer formation between the $H_B= 0$ case, where $KE \simeq 2.4$ and the $H_B = 0.1$ case, where $KE \simeq 1.8$). 
Crucially, we also find that layers 
take much longer to appear in the magnetic cases than for $H_B = 0$, and that the first layered configuration has fewer layers (see Section \ref{sec:gammanum3}). For $H_B = 0.03$ for instance, layered convection starts around $t \simeq 4300$ with four layers which rapidly merge into three layers around $t \simeq 4500$. By $t \simeq 5000$, they have merged down to two layers and by the end of the simulation are in the process of merging into a single layer. In the case of $H_{B} = 0.1$, the system stays in a state of weak wave-like convection for a very long time before layers start to emerge. Layered convection does not start until around $t \simeq 8300$, and again begins  in a four-layer phase. It is interesting to note that the rate at which the layers merge, however, appears to be relatively independent of $H_B$.



Finally, we also see that the kinetic energy in the layered phase depends on the magnetic field strength. Indeed, the black horizontal lines in Figure \ref{fig:energies} indicate, from top to bottom, the average kinetic energy of the two, three and four layered phases in the non-magnetic ($H_B = 0$) case, and help guide the eye for comparison with the magnetic cases. We see that the kinetic energy in each of the layered phases are lower for $H_B = 0.03$ and $H_B = 0.1$ than in the non-magnetic case, and that the effect becomes stronger as the magnetic field increases. This suggests that a strong enough magnetic field can substantially reduce the efficiency of layered convection. Meanwhile, we also see that the magnetic energy increases not only with $H_B$ (as expected), but also with each merger during the layered phase, demonstrating that some of the kinetic energy from the convective motions is being converted into magnetic energy.

In summary, our DNS demonstrate  that the presence of a vertical magnetic field can both slow down (or suppress, in some cases shown in the next sections) the formation of layers, and reduce the kinetic energy of the flow in layered convection. Given the likely importance of layer formation and mergers in ODDC near the core of intermediate-mass stars \citep[see Section \ref{sec:intro} and][]{MooreGaraud2016}, it is therefore crucial to establish whether these statements still hold at stellar parameter values (which would have significant consequences for stellar evolution). To do so, we first need to determine how the layer-forming $\gamma$-instability is affected by the magnetic field, both qualitatively and quantitatively. The next section investigates this question in detail. The reader merely interested in the answer may skip to Section \ref{sec:ccl}.

\section{Layer formation}
\label{sec:gamma}

\subsection{Theory}
\label{sec:gammatheor}

As mentioned in Section \ref{sec:intro}, in the absence of magnetic fields, the spontaneous emergence of convective layers from homogeneous ODDC has been attributed to the so-called $\gamma$-instability  \citep{radko2003mechanism,rosenblumal2011}. 
We now demonstrate that a large-scale uniform magnetic field does not {\it directly} influence the $\gamma$-instability mechanism, so the latter is still expected to operate as before. 

While most standard fluid instabilities are instabilities of the original Navier-Stokes equations, and describe the evolution of perturbations from an initial laminar state, the $\gamma$-instability is an instability of the Reynolds-averaged equations (where in this case, the average is simply a horizontal average) which model the evolution of large-scale spatial modulations to a pre-existing turbulent state. It is caused by a positive feedback loop between the vertical turbulent temperature and compositional fluxes (that cause an evolution of the temperature and composition gradients) and the local stratification (that controls these turbulent fluxes), which is schematically illustrated, for instance, in Fig.~3 of \citet{Garaud2018}.
The physical mechanism driving the $\gamma$-instability is quite subtle. While we describe the mathematical details below, readers interested in a more schematic illustration are referred to Fig.~3 of \citet{Garaud2018}.

This feedback loop can be studied as follows \citep{radko2003mechanism,Mirouh2012}. First, we take the horizontal average of (\ref{eq:tempeq}) and (\ref{eq:Ceq}), which results in:   
\begin{eqnarray}
    \frac{\partial \bar T}{\partial t } =  \frac{\partial^2 \bar T}{\partial z^2} - \frac{\partial F_T}{\partial z} = - \frac{\partial F^{tot}_T}{\partial z}, \label{eq:Tflux} \\
        \frac{\partial \bar C}{\partial t } =  \tau \frac{\partial^2 \bar C}{\partial z^2} -\frac{\partial F_C}{\partial z} = - \frac{\partial F^{tot}_C}{\partial z} \label{eq:Cflux}, 
\end{eqnarray}
where $\bar T$ is the horizontal average of the fluctuations $\hat T$, and similarly for $\bar C$. We have introduced the turbulent temperature and composition fluxes $F_T = \overline{\hat u_z \hat T}$ and $F_C = \overline{\hat u_z \hat C}$, as well as the total temperature and composition fluxes 
\begin{eqnarray}
F^{tot}_T = 1 - \frac{\partial \bar T}{\partial z} + F_T,  \\
F^{tot}_C = \tau R_0^{-1} - \tau \frac{d\bar C}{dz} + F_C. 
\end{eqnarray} 
Equations (\ref{eq:Tflux}) and (\ref{eq:Cflux}) describe the first part of the feedback loop, namely how spatial modulations of the fluxes cause a temporal evolution of the temperature and composition stratification. Note how the magnetic field does not appear in their derivation explicitly -- in fact, the same equations apply in the hydrodynamic case. However, it does so implicitly since the turbulence, and therefore the turbulent fluxes, are affected by the field. 

In a strictly homogeneous turbulent system, $\bar T \simeq \bar C \simeq 0$, and the temperature and composition gradients are constant and equal to the non-dimensional background values of $-1$ and $-R_0^{-1}$, respectively. In that case the temperature and composition fluxes $F^{tot}_T$ and $F^{tot}_C$ are independent of height ($z$) in the domain. This is the state illustrated in the second row of Figure \ref{fig:fluxes_layers}. 

A crucial ingredient for the second part of the feedback loop in Radko's $\gamma$-instability theory is the fact that, for a given fluid (i.e. at fixed $Pr$, $\tau$ and $D_B$) and a given background magnetic field (i.e. fixed $H_B$), the intensity and properties of double-diffusive turbulence depend on the local temperature and composition gradients only via the local inverse density ratio \citep{radko2003mechanism}, written by \citet{Mirouh2012} for ODDC as
\begin{equation}
R^{-1} = \frac{R_0^{-1} - d\bar C/dz}{1 - d\bar T/dz}.
\label{eq:localR}
\end{equation}
This quantity is the ratio of the local density gradient due to composition stratification, to the local density gradient due to temperature stratification, and is equal to $R_0^{-1}$ in the homogeneous state.

To model the $\gamma$-instability mathematically, \citet{radko2003mechanism} then introduced two important non-dimensional quantities: the thermal Nusselt number, which is the ratio of the total temperature flux to the diffusive temperature flux
\begin{equation}
    Nu_T = \frac{F_T^{tot} }{1 - d\bar T/dz},
    \label{eq:NuT}
\end{equation}
and the (inverse) flux ratio 
\begin{equation}
    \gamma_{tot}^{-1} = \frac{F_C^{tot} }{F_T^{tot}},
    \label{eq:gammatotinv}
\end{equation}
which is the ratio of the total composition flux to the total temperature flux. The key assumptions discussed above can be expressed mathematically by requiring that both $Nu_T$ and $\gamma_{tot}^{-1}$ should only be functions of $R^{-1}$ (at fixed $Pr$, $\tau$, $D_B$  and $H_B$). With this assumption, we have 
\begin{equation}
    F_T^{tot} = \left(1 - \frac{d\bar T}{dz}\right) Nu_T(R^{-1}),  \mbox{ and }  F_C^{tot} = \gamma_{tot}^{-1}(R^{-1}) F_T^{tot}.      \label{eq:fluxes}
\end{equation}




If the functions $Nu_T(R^{-1})$ and $\gamma_{tot}^{-1}(R^{-1})$ are known, then equations (\ref{eq:Tflux}), (\ref{eq:Cflux}), (\ref{eq:localR}), and (\ref{eq:fluxes}) form a closed system describing the evolution of large-scale spatial inhomogeneities $\bar T(z,t)$ and $\bar C(z,t)$. These equations can then be linearized around the previously introduced homogeneous turbulent state (which has $\bar T = \bar C =0$, and $R^{-1} = R_0^{-1}$), in the limit where $\bar T$ and $\bar C$ are very small, to demonstrate that perturbations of the kind $\bar T(z,t) \sim e^{i k z + \Lambda t}$ and $\bar C(z,t) \sim e^{i k z + \Lambda t}$ grow or decay exponentially with time, with a growth rate 
\begin{equation}
    \Lambda = \sigma k^2,
\end{equation}
where $\sigma$ satisfies the quadratic equation 
\begin{equation}
    \sigma^2 + a \sigma + b  = 0 ,
    \label{eq:Lambdaquad}
\end{equation}
and where the coefficients $a$ and $b$ depend on the properties of the homogeneous state as
\begin{eqnarray}
    && a = A_{Nu}(1- R_0 \gamma_0^{-1}) + Nu_0 (1- A_\gamma R_0) , \nonumber \\
    && b = - A_\gamma Nu_0^2 R_0 ,
    \label{eq:abeq}
\end{eqnarray}
with 
\begin{eqnarray}
    Nu_0 = Nu_T(R_0^{-1}), \quad \gamma^{-1}_0 = \gamma_{tot}^{-1} (R_0^{-1}), \nonumber \\ 
    A_{Nu} = -R_0^{-1}\left. \frac{d Nu}{d R^{-1}}  \right|_{R_0^{-1}}, \quad A_{\gamma} = -R_0^{-1}\left. \frac{d \gamma^{-1}_{tot}}{d R^{-1}}  \right|_{R_0^{-1}}.
    \label{eq:coeffdef}
\end{eqnarray}
A detailed derivation of this result is presented in \citet{rosenblumal2011} and \citet{Mirouh2012}. Crucially, we see from the steps outlined above that the presence of a magnetic field does not explicitly appear in this theory. However it does so implicitly by influencing the functions $Nu_T(R^{-1})$ and $\gamma_{tot}^{-1}(R^{-1})$, and therefore  the coefficients $a$ and $b$ of the quadratic equation (\ref{eq:Lambdaquad}).   

Equation (\ref{eq:Lambdaquad}) immediately shows that a necessary condition for instability, i.e the existence of solutions with $\sigma > 0$, is  
\begin{equation}
b < 0 \rightarrow \frac{d \gamma_{tot}^{-1}}{dR^{-1}} < 0 ,
\label{eq:gammacriterion}
\end{equation}
i.e. that the inverse flux ratio should be a decreasing function of the inverse density ratio \citep{radko2003mechanism,Mirouh2012}. 
This $\gamma$-instability theory,  and more specifically equation (\ref{eq:Lambdaquad}) was found to correctly predict the growth rate of large-scale perturbations in both fingering convection in the ocean \citep{radko2003mechanism,Stellmach2011} and ODDC in stars \citep{rosenblumal2011,Mirouh2012}.  We now investigate whether this remains true for magnetized ODDC. 

\subsection{Comparison with numerical experiments 1: layering vs. no layering}
\label{sec:gammanum1}

To test the $\gamma$-instability theory derived above, we follow the same steps as in \citet{Mirouh2012}. The first step consists in measuring the functions $Nu_T(R^{-1})$ and $\gamma_{tot}^{-1}(R^{-1})$ in homogeneous ODDC turbulence at fixed values of $Pr$, $\tau$, $D_B$ and $H_B$. To do so, we run and analyze a number of small-domain DNS, each of which is performed in a triply-periodic cube of size $100d \times 100d \times 100d$ with a resolution of $384 \times 384 \times 384$ equivalent grid points. Each simulation is initialized as those of Section \ref{sec:num}, with all fields set to zero except for $\hat T$ and $\hat C$ which are seeded with small random noise, and $\hat B_z$ which is set to 1. The equations are  evolved either until the first set of layers form, or until we have acquired enough data to be certain that they do not. As in \citet{Mirouh2012}, we use the fact that
\begin{equation}
    \langle \hat u_z \hat T \rangle \simeq \langle | \nabla \hat T |^2 \rangle \quad \mbox{ and }
 \quad         \langle \hat u_z \hat C \rangle \simeq R_0 \tau\langle | \nabla \hat C |^2 \rangle
\end{equation}
when the flow is in a statistically stationary state, to compute 
\begin{eqnarray}
    Nu_T =  F_T^{tot} = 1 + \langle | \nabla \hat T |^2 \rangle,  \\ 
    \gamma_{tot}^{-1} = \frac{F_C^{tot}}{F_T^{tot}} = \frac{\tau R_0^{-1} + R_0 \tau\langle | \nabla \hat C |^2\rangle}{1 + \langle | \nabla \hat T |^2 \rangle}. 
\end{eqnarray}
We then take a time average of $Nu_T$ and $\gamma_{tot}^{-1}$ in the statistically stationary homogeneous ODDC phase. 

Note that our determination of the appropriate time interval for this average deviates somewhat from the method used by \citet{Mirouh2012} (see their Appendix B), mostly for reasons of simplicity (see below), but also because their method turns out to be somewhat inconsistent with the assumptions of the $\gamma$-instability theory outlined above. In what follows, we use the same steps as \citet{Mirouh2012} to determine the averaging start time, but set the averaging end time to be either when the horizontally-averaged density profile begins to deviate substantially from the background linear profile (when layers form), or at the end of the simulation (when no layers form).  \citet{Mirouh2012}, by contrast, stopped the averaging either when fully convective layers appear, or when large-scale gravity waves begin to dominate the simulation, whichever happens the soonest. As a result, their estimates for the convective fluxes are systematically higher than ours in very low $R_0^{-1}$ simulations where layers form early, because their average includes times during which the layers have almost (but not completely) overturned, while we stop before this happens. Meanwhile, their estimates for the convective fluxes are often lower than ours at moderate and high $R_0^{-1}$, because the large-scale gravity waves that eventually appear in that limit \citep[see][]{Mirouh2012,Moll2016}, which they discard but we keep, cause a non-negligible amount of transport. 

\begin{figure}
\includegraphics[width = \textwidth]{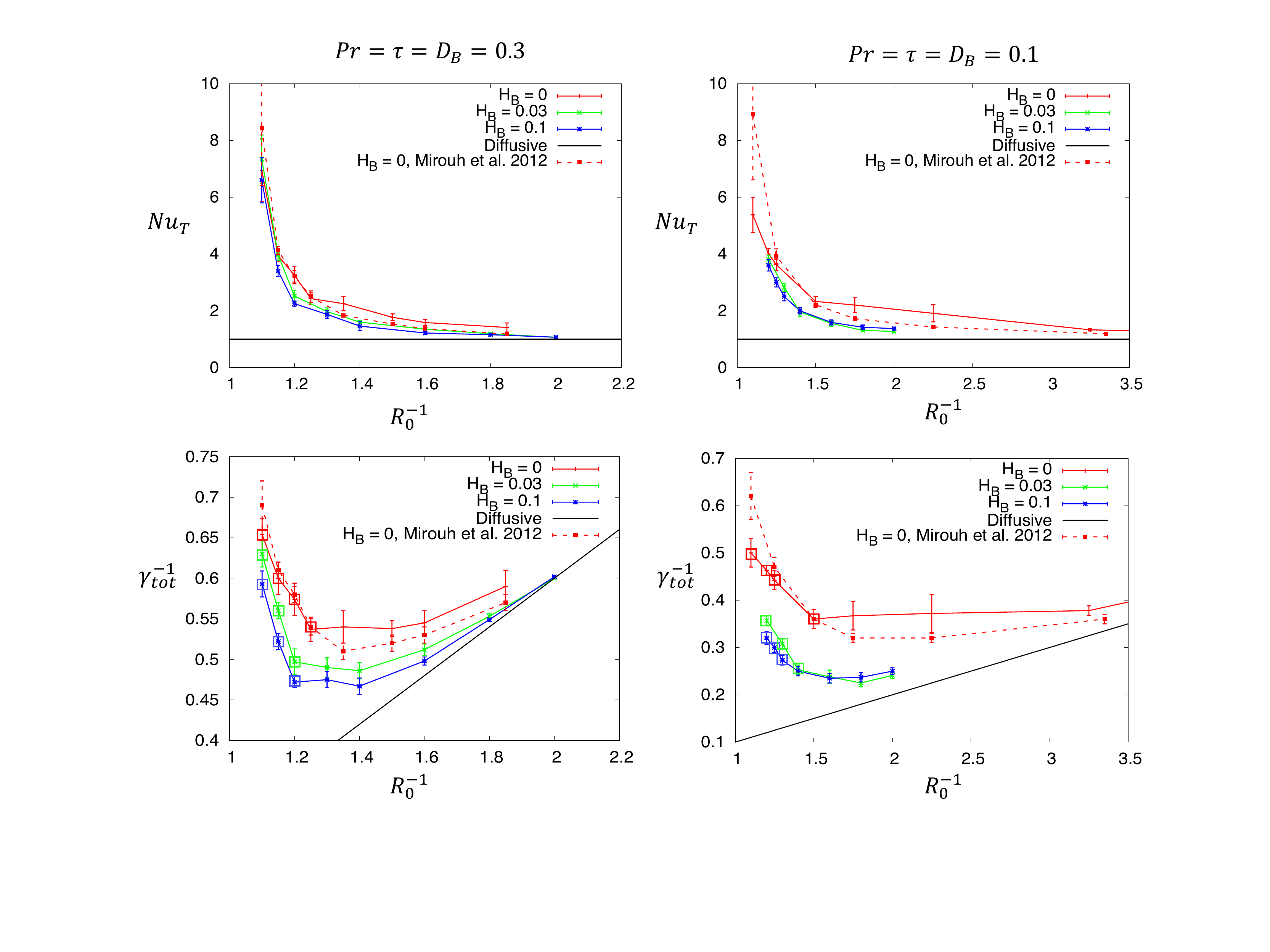}
\caption{In each panel, the solid curves show the functions $Nu_T(R_0^{-1})$ (top) and $\gamma_{tot}^{-1}(R_0^{-1})$ (bottom) extracted from small-domain DNS as described in the main text, for various input parameters. The dashed curves show the same quantities extracted using the method in \citet{Mirouh2012}. The left column has $Pr = \tau = D_B = 0.3$, the right column has  $Pr = \tau = D_B = 0.1$. Various magnetic field strengths are presented, as described in the legend. Symbols that are surrounded by an open square in the bottom row denote simulations for which layers eventually appear. The black line in each panel shows the values of $Nu_T = 1$ and $\gamma_{tot}^{-1} = \tau R_0^{-1}$ corresponding to the diffusive solution (when $\langle \hat u_z \hat T\rangle = \langle \hat u_z \hat C\rangle  =0$). All curves asymptote to this solution as $R_0^{-1}$ approaches the threshold $R_c^{-1} = (\Pr+1)/(\Pr+\tau)$. }
\label{fig:Nugamma}
\end{figure}

Figure \ref{fig:Nugamma} shows as solid curves the functions $Nu_T(R_0^{-1})$ (top) and $\gamma_{tot}^{-1}(R_0^{-1})$ (bottom) extracted using this new simplified method, for $Pr = \tau = D_B = 0.3$ (left) and $Pr = \tau = D_B = 0.1$ (right), for three values of the field strength ($H_B = 0$, $H_B = 0.03$ and $H_B = 0.1$). The $H_B = 0$ data (red solid curve) is a re-analysis of the hydrodynamic simulations from \citet{Mirouh2012} using the new method, and the red dashed line shows, for comparison, the values of  $Nu_T(R_0^{-1})$ and $\gamma_{tot}^{-1}(R_0^{-1})$ reported in their Table 5, which they had extracted using their method. The difference between the solid and dashed red curves illustrates the fairly substantial impact of choosing a different interval for the time averages, as mentioned above. That impact is, however, smaller than the impact of adding a magnetic field.

 Indeed, and consistent with the results briefly discussed in Section \ref{sec:num}, we see that even a weak magnetic field (e.g. $H_B = 0.03$, green curves) can sometimes substantially reduce the turbulent heat flux, decreasing $Nu_T$. This in turn changes the shape of the $\gamma_{tot}^{-1}(R_0^{-1})$ curve, which gradually tends towards the diffusive flux ratio $\gamma_{diff}^{-1}(R_0^{-1}) = \tau R_0^{-1}$ (solid black lines). The theory developed in the previous section, combined with these results, therefore shows that a large-scale magnetic field has an {\it indirect} impact on layer formation, by affecting $Nu_T$ and $\gamma_{tot}^{-1}$, which changes the values of the coefficients of the quadratic (\ref{eq:Lambdaquad}), and in turn, the $\gamma$-instability growth rate.

 A rapid, qualitative way of testing the predictions of the $\gamma$-instability theory is to verify that all simulations that were run at parameters for which $d \gamma_{tot}^{-1}/ dR_0^{-1} < 0$ (cf. equation \ref{eq:gammacriterion}) eventually transition into layered convection. This is indeed almost always the case. In Figure \ref{fig:Nugamma} (bottom row), simulations that ultimately become layered have an additional open square surrounding the original symbol. We see that the square is present whenever $\gamma_{tot}^{-1}$ is a rapidly decreasing function of $R_0^{-1}$, as in \citet{Mirouh2012}. We also see that no layers ever form when $\gamma_{tot}^{-1}$ increases with $R_0^{-1}$, consistent with the theory. The only simulations in which the data disagrees with the theory are those for which  $\gamma_{tot}^{-1}$ decreases very slowly with $R_0^{-1}$ (near the minimum of the curve), where layering was not observed even though the $\gamma$-instability ought to be active. This is possibly because the growth rate of the $\gamma$-instability is too low in that limit, and other effects that are not accounted for (such as the presence of large-scale gravity waves in the system) further damp it \citep[see, e.g.][for related effects in oceanic fingering convection]{Traxler2011b}.

Notwithstanding this minor discrepancy, we now also see that a sufficiently strong magnetic field can, in some cases, completely suppress layer formation. Indeed, consider the case with $Pr = \tau = D_B = 0.1$. At these parameters, Figure \ref{fig:Nugamma} (bottom right panel) shows that layers form at $R_0^{-1} = 1.4$  in the non-magnetic case, and in the $H_B = 0.03$ case, but not for $H_B = 0.1$. At the same time it provides a tentative explanation why, namely by moving the minimum of the  $\gamma_{tot}^{-1}(R_0^{-1})$ curve slightly to the left, which eventually stabilizes the system to the $\gamma$-instability at these parameters.  

\subsection{Comparison with numerical experiments 2: growth rate of the layering modes in small domains}
\label{sec:gammanum2}

Even when layers eventually form, we have seen in Section \ref{sec:num} that a weak magnetic field can  significantly delay the onset of layered convection. The same was found to be true in all of the small domain simulations discussed in Section \ref{sec:gammanum1}. The theory presented in Section \ref{sec:gammatheor} suggests a possible explanation for this delay, namely that the magnetic field reduces the turbulent temperature and composition fluxes, which in turn reduces the growth rate of the $\gamma$-instability. We now test this idea more quantitatively.

We begin by comparing the $\gamma$-instability theory predictions against data from small-domain simulations. This turns out to be easier than starting with the large-domain simulations of Section \ref{sec:num}, because the latter have many modes growing simultaneously (layering modes and sometimes large-scale gravity waves). Their presence can obfuscate the dynamics of the $\gamma$-instability, as discussed below. We therefore focus on four available small-domain layer-forming simulations, whose parameters are presented in 
Table \ref{tab:coeffs_small}. The table also shows the extracted values of $Nu_T$, $\gamma_{tot}^{-1}$, $A_{Nu}$ and $A_\gamma$ for these simulations. The derivative terms $A_{Nu}$ and $A_\gamma$ are computed using second-order finite differences using simulations at values of $R_0^{-1}$ on both sides of the target one. With this information, we can then evaluate the quadratic coefficients $a$ and $b$ using (\ref{eq:abeq}), and solve (\ref{eq:Lambdaquad}) for $\sigma$ (also shown in Table \ref{tab:coeffs_small}). Finally, to compute the growth rate $\Lambda_n$ of a particular layering "mode" with $n$ layers, we use the fact that 
\begin{equation}
    \Lambda_n = \sigma k_n^2 \mbox{   with  } k_n = \frac{2 n \pi}{L_z}.
    \label{eq:lambda_n}
\end{equation} 

\begin{table}[]
    \centering
    \begin{tabular}{c|c|c|c|c|c|c|c}
       $Pr=\tau=D_B$ & $R_0^{-1}$ & $H_B$ & $Nu_T$ & $\gamma_{tot}^{-1}$ & $A_{Nu}$ & $A_{\gamma}$ & $\sigma$   \\
        \hline

 0.3 & 1.15  & 0.03   &  $2.97 \pm 0.20$   &  $0.56 \pm 0.01$              &        $55.4$    &      $1.53$         &  $0.75$\\ 
 0.3  & 1.15 & 0.1 &  $2.4 \pm 0.2$   & $0.52 \pm 0.01$               &    $50$       &   $1.39$           & $0.51$ \\
 \hline
0.1  & 1.3 & 0.03 &  $1.94 \pm 0.2$   & $0.31 \pm 0.01$               &    $12.7$       &   $0.29$           & $0.40$ \\
0.1  & 1.3 & 0.1 &  $1.5 \pm 0.15$   & $0.27 \pm 0.01$               &    $10.7$       &   $0.57$           & $0.27$\\
\hline
     \end{tabular}
    \caption{Values of $Nu_T$, $\gamma_{tot}^{-1}$, $A_{Nu}$, $A_{\gamma}$ extracted from DNS at $R_0^{-1} = 1.15$, $Pr = \tau = D_B = 0.3$ (top two lines) and $R_0^{-1} = 1.3$, $Pr = \tau = D_B = 0.1$ (bottom two lines). The derivatives $A_{Nu}$ and $A_\gamma$ are computed using finite differencing with data at neighboring values of $R_0^{-1}$.  }
    \label{tab:coeffs_small}
\end{table}

We immediately see from Table \ref{tab:coeffs_small} that $\sigma$ decreases substantially when $H_B$ increases from 0.03 to 0.1 with all other parameters fixed. This confirms our above hypothesis that the magnetic field can reduce the growth rate of layers. 

\begin{figure}[h!]
\includegraphics[width = 0.9\textwidth]{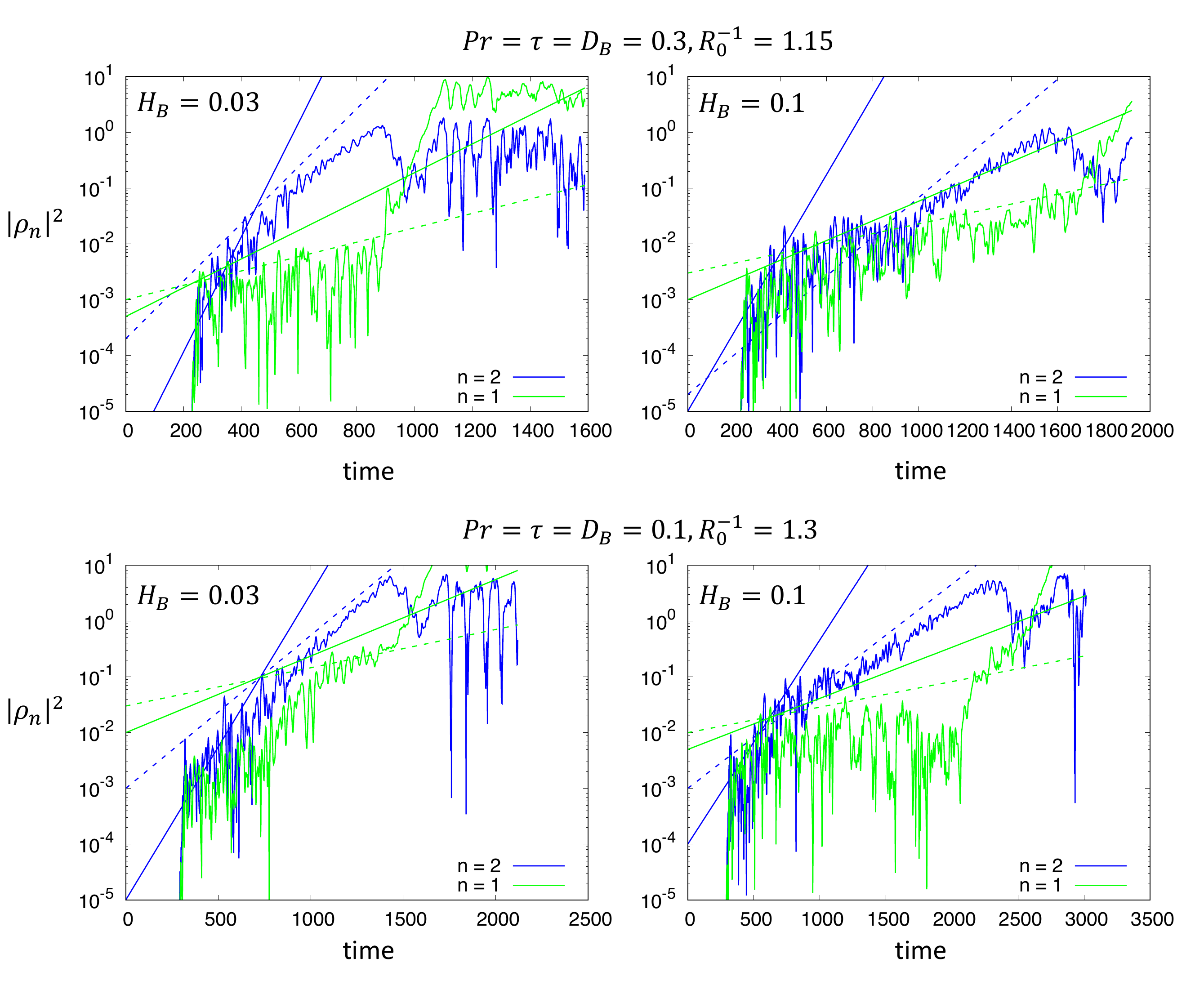}
\caption{Density spectral power of layering modes as a function time, for the 4 simulations analyzed in Table \ref{tab:coeffs_small} (see text for detail). In each panel, the colored curve show $|\rho_n(t)|^2$ for $n = 2$ (blue) and $n=1$ (green), and the solid line of the same color shows the predicted exponential growth of the mode according to the $\gamma$-instability theory. The dashed line of the same color shows the same with half the growth rate, which appears to fit the data much better overall. }
\label{fig:testgamma}
\end{figure}

For a more direct test of the $\gamma$-instability theory against the data, we now compare the growth rates of the layering modes observed in the simulations to those predicted by the theory. We extract the time-dependent amplitude of these modes from the DNS by computing the Fourier expansion of the horizontally-averaged density perturbations, namely 
\begin{equation}
    \bar \rho(z,t) = - \bar T(z,t) + \bar C(z,t) = \sum_{n} \rho_n(t) e^{ik_n z}.
\end{equation}
The quantity $|\rho_n|^2(t)$, called the density spectral power hereafter, is plotted in Figure \ref{fig:testgamma} for the four simulations presented in Table \ref{tab:coeffs_small}, for modes leading to $n = \{ 1,2 \}$ layers in each case. Modes with $n > 2$ are not usually found to grow in small domain simulations. 

The $\gamma$-instability theory predicts that $|\rho_n|^2(t) \propto e^{2\Lambda_n t}$, with $\Lambda_n$ given by (\ref{eq:lambda_n}) for a given Fourier mode (equivalently, number of layers) $n$.  
We compare these predictions (colored solid lines) to the data for each mode in Figure \ref{fig:testgamma}. We see that in all cases, the model is appropriate for the $n=2$ mode at very early times, but overestimates its growth rate by a factor of about two at later times (after $t \sim 600$). Predictions made with half the growth rate (dashed lines) appear to fit the data better then. The situation is not as clear for the $n=1$ mode (in some cases, the above statements hold, and in some others, they do not), which is perhaps not too surprising because the latter is growing intrinsically slowly, in a fairly turbulent environment.

The fact that some layering modes grow slower than expected at later times can already be seen in some of the hydrodynamic simulations of \citet{Mirouh2012} but was not discussed in that paper. However, we now see that this is a relatively systematic effect. Inspection of the total density profile before, during, and after the time where the mode growth rate starts decreasing reveals that this corresponds to the point where the total density profile is no longer linear. At this point, the linearization procedure that leads to the derivation of the quadratic growth rate equation (\ref{eq:Lambdaquad}) is no longer valid, and it is therefore not surprising to see that the model no longer fits the data at the quantitative level. We do see, however, that the mode continues to grow, albeit at a smaller rate.  

\subsection{Comparison with numerical experiments 3: growth rate of the layering modes in large domains}
\label{sec:gammanum3}

\begin{figure}[h!]
\includegraphics[width = 0.9\textwidth]{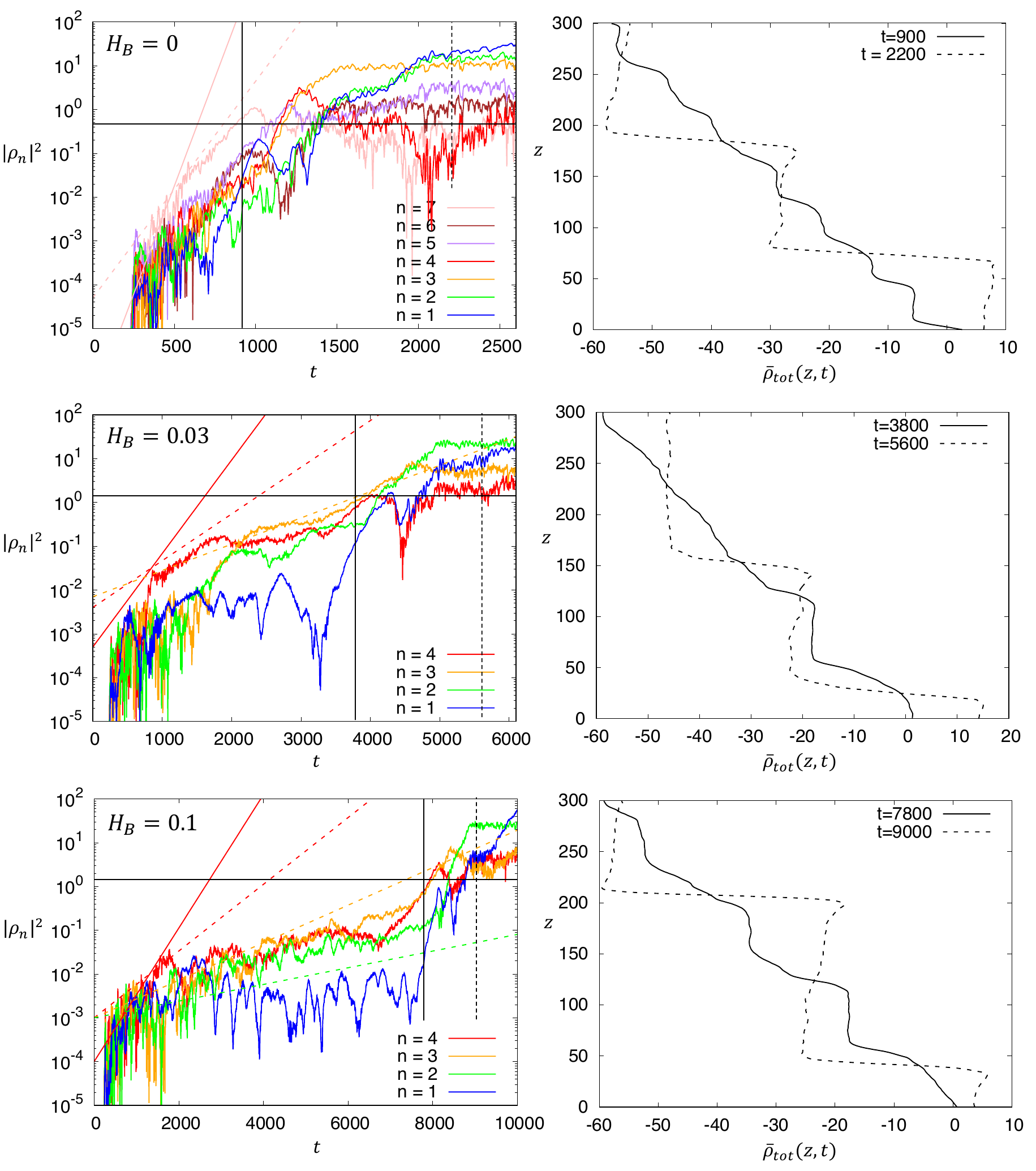}
\caption{Left: Density spectral power $|\rho_n|^2$ as a function of time for various layering modes for the simulations discussed in Section \ref{sec:num} with $Pr = \tau = D_B = 0.3$, and $R_0^{-1} = 1.2$. The magnetic field increases from top to bottom. The matching colored solid lines show the predicted growth of the layering mode according to the $\gamma$-instability theory, and the dashed colored lines shows the same with half that growth rate. The horizontal black line shows the overturning threshold for the mode with $n_L$ layers, with $n_L = 7$ in the non-magnetic case (top), and $n_L=4$ in the other cases (middle and bottom). The solid and dashed vertical lines mark the times at which density profiles are shown (right). The solid vertical line also marks the time at which the kinetic energy first increases in Figure \ref{fig:energies}. Right: Horizontally averaged total density profiles shown at selected times in the same simulations.}
\label{fig:DenzSpec}
\end{figure}

To conclude this analysis, we now compare the predictions of the $\gamma$-instability theory with the layering mode data for the three large-domain simulations presented in Section \ref{sec:num}, which have $Pr = \tau = D_B = 0.3$, $R_0^{-1} = 1.2$. The results are presented in Figure \ref{fig:DenzSpec}. Each row of Figure \ref{fig:DenzSpec} corresponds to a different magnetic field strength, from top to bottom $H_B = 0$, $0.03$ and $0.1$. The density spectral power of the first $n_L$ layering modes are shown in each case, where $n_L$ is the number of layers of the mode that dominates at early times. As discussed in Section \ref{sec:num}, $n_L = 7$ for $H_B = 0$ (pink curve), while $n_L = 4$ for $H_B = 0.03$ and $H_B = 0.1$ (red curve). Modes with $n>n_L$ (not shown) are not found to grow in the simulation. It is worth noting that it is not clear why $n_L$ is smaller in the magnetized cases than in the hydrodynamic case, and whether this is a systematic result, or a coincidence. We discuss this issue in Section \ref{sec:ccl}.  

Table \ref{tab:coeffs} shows $Nu_T$, $\gamma_{tot}^{-1}$, $A_{Nu}$,  $A_\gamma$ and $\sigma$ extracted from the small domain simulations, this time for the parameter values used here. All the quantities are computed as in  Table \ref{tab:coeffs_small}. The predicted growth rates, calculated as in the previous Section for each dominant mode (i.e. a mode whose amplitude is much larger than the others), are shown in Figure \ref{fig:DenzSpec} as solid lines. Dashed lines show how these modes would grow at half the predicted growth rates. As before, we find that most modes grow at the predicted rate at early times (before $t \sim 600$), but then later continue at about half the predicted rate. The strongest field case ($H_B = 0.1$) is a little different, however, as discussed below.

\begin{table}[]
    \centering
    \begin{tabular}{c|c|c|c|c|c}
        $H_B$ & $Nu_T$ & $\gamma_{tot}^{-1}$ & $A_{Nu}$ & $A_{\gamma}$ & $\sigma$   \\
        \hline
     0  &  $3.28 \pm 0.15$   &  $0.575 \pm 0.01$
    &  $ 19.3$ &   $0.72 $ & $0.53$ \\ 
        0.03  &  $2.57 \pm 0.20$   &  $0.50 \pm 0.01$         
        &   $25.3$    & $0.99 $ &  $ 0.35 $\\ 
        0.1 &  $2.27 \pm 0.09$   & $0.475 \pm 0.01$              
        &    $19.7 $ &  $ 0.76$  & $0.25 $
    \end{tabular}
    \caption{Values of $Nu_T$, $\gamma_{tot}^{-1}$, $A_{Nu}$, $A_{\gamma}$ extracted from DNS at $R_0^{-1} = 1.2$, $Pr = \tau = D_B = 0.3$. The derivatives $A_{Nu}$ and $A_\gamma$ are computed using finite differencing with data at nearby values of $R_0^{-1}$. 
    }
    \label{tab:coeffs}
\end{table}

For the non-magnetic case, we see that the 7-layer mode clearly dominates at early times until $t \simeq 1000$. It grows roughly at the predicted rate until $t \simeq 600$, at which point it begins to grow at about half the predicted rate. Around $t \simeq 900$, we saw in Figure \ref{fig:energies} that the kinetic energy of the flow increases substantially. Inspection of the total density profile at that time (see right-hand panel, solid black line) shows that it is no longer linear, but instead, clearly exhibits the presence of a 7-layer mode, with some of the layers already being fully convective (i.e. with a density that increases with height). It is therefore not surprising to see that the mode stops growing shortly after $t = 900$. 

Note that the amplitude above which a {\it single} mode with $n$ layers causes an inversion in the total density profile $\bar \rho_{tot}(z,t) = (1-R_0^{-1})z + \bar \rho(z,t)$ was given by \citet{rosenblumal2011} to be 
\begin{equation}
    |\rho_n|_{crit} =  \left|\frac{R_0^{-1} - 1}{2k_n}\right|.
\end{equation}
This threshold, computed for $n = n_L$, is shown as a horizontal black line in each panel of Figure \ref{fig:DenzSpec}. We see that in the non-magnetic calculation, the time at which the kinetic energy of the flow begins to increase ($t=900$, vertical black line) corresponds to the time at which the $7$-layer mode amplitude reaches $|\rho_n|_{crit}$. Beyond that point, and as discussed in Section \ref{sec:num}, convective layers appear and rapidly begin to merge. We see, accordingly, that the dominant mode changes with time. A second density profile is shown at $t = 2200$ (black dashed line), showing two convective layers separated by thin interfaces. 

In the weak magnetic field case (middle row) the situation is overall similar -- a dominant layering mode  grows from the $\gamma$-instability, and layers eventually appear. This mode has fewer layers ($n_L = 4$) than in the non-magnetic case, but its growth rate continues to be reasonably well predicted using half the theoretical growth rate after $t = 600$. By contrast with the non-magnetic case, however, it does not remain dominant until it overturns, but instead, appears to somehow aid the growth of larger-scale layering modes ($n=3$ and $n=2$) around $t = 1800$.  Inspection of the density profile at that time reveals the presence of a single, shallow convective layer in the lower half of the domain, that may have been created earlier than expected by a large-scale gravity wave breaking in a region of the domain whose stratification was already weakened by the presence of the layering modes. This convective layer remains in place for the rest of the simulation, and therefore couples the $n=2$, $n=3$ and $n=4$ layering modes. All three modes then continue to grow at a slower rate, until the convective layers are fully established around $t = 3800$ (see right-side panel with two layers, one centered around $z = 80$ and another centered around $z = 0$.)

Finally, the situation for the stronger field case $H_B = 0.1$ is once again a little different. The $n_L = 4$ mode is dominant at early times, but seems to couple with both $n = 3$ and $n=2$ modes around $t = 2000$. By contrast with the $H_B = 0.03$ simulation, however, there is no evidence for a convective layer at that point, and the total density profile remains close to being linear until $t \simeq 6000$. All three modes nevertheless continue to grow at a rate that is consistent with half the predicted growth rate of the $n=2$ mode. Overturning convection is again triggered a little earlier than the time at which the amplitude of any of the modes reaches the critical threshold $|\rho_n|_{crit}$.

\section{Summary and discussion}
\label{sec:ccl}

In the previous section, we have demonstrated that the $\gamma$-instability theory of \citet{radko2003mechanism} continues to be a good model for layer formation in magnetized ODDC, at least qualitatively, correctly predicting whether layers form or not. It can sometimes overestimate the growth rate of layer-forming modes by a factor of order unity, but this is not too much of a concern. Indeed, being driven by turbulent mixing processes, the timescale for layer formation is much shorter than any evolutionary timescale (regardless of the magnetic field strength). This suggests that layers would appear almost instantaneously from the perspective of stellar evolution whenever the $\gamma$-instability is excited. We have also seen that layers rapidly merge once they form, ultimately leading (in stars) to a fully-mixed convective layer. This merger process does not appear to be affected by the magnetic field, at least for the parameters achievable in the DNS. 

These results, when combined, suggest that there are two possible outcomes for ODDC-unstable regions in stars: either the $\gamma$-instability is excited (i.e. has a positive growth rate), in which case the region rapidly becomes fully convective, or the $\gamma$-instability is not excited (i.e. has a negative growth rate), in which case the region remains in a state of weakly turbulent ODDC. This conclusion is not new \citep{MooreGaraud2016}, but we now confirm that it remains true for magnetized ODDC.   

In Section \ref{sec:gammatheor}, we recalled that a sufficient criterion for the  $\gamma$-instability to occur is that the inverse flux ratio $\gamma_{tot}^{-1}$ (i.e. the ratio of the total composition flux to the total temperature flux caused by the basic ODDC instability) should be a decreasing function of the inverse density ratio $R_0^{-1}$. This criterion still applies in magnetized ODDC. But in Section \ref{sec:gammanum2} we also found that even a relatively weak magnetic field can change the shape of the $\gamma_{tot}^{-1}(R_0^{-1})$ curve and move the position of its minimum, thereby shrinking the region of parameter space unstable to layering. As a result, systems whose stratification is unstable to the $\gamma$-instability in the absence of magnetic fields can be stabilized when the field exceeds a certain threshold. 

Unfortunately, the threshold magnetic field that was relevant to our DNS is not directly applicable to stars. This is because stellar fluids generally have much smaller diffusivity ratios $Pr$, $\tau$, and $D_B$ than what can be achieved numerically, which means that we cannot directly compute the $\gamma_{tot}^{-1}(R_0^{-1})$ curve relevant for stars (at least, not with currently available supercomputing power). Future theoretical work will therefore need to identify the mechanism responsible for the saturation of magnetized ODDC, to better predict the dependence of the turbulent fluxes on all input parameters, especially the field strength $H_B$. With that information, we might then be able to predict the shape of   $\gamma_{tot}^{-1}(R_0^{-1})$ at stellar values of $Pr$, $\tau$ and $D_B$, for varying $H_B$ \citep[similar to the model of][for the non-magnetic case]{Mirouh2012}. The position of the minimum of that curve, and its dependence on the magnetic field strength, then determines whether a particular ODDC-unstable region in the star, with a given stratification characterized by $R_0^{-1}$, is unstable to layering.

In the event the new model reveals that ODDC in stars can indeed be stabilized against the $\gamma$-instability by a sufficiently strong (but still realistic) magnetic field, this could lead to interesting observational predictions. Indeed, \citet{MooreGaraud2016} showed the ODDC-unstable region surrounding the Ledoux-sized core of intermediate-mass stars rapidly becomes fully convective (as a result of the $\gamma$-instability) in the non-magnetic case. These stars therefore have a larger-than-expected convective core whose size is appropriately computed using the Schwarzschild criterion instead.
Intermediate-mass stars with a sufficiently strong magnetic field would, by contrast, remain in a state that has a smaller convective core, surrounded by a region of weak ODDC. Asteroseismic observations of Ledoux-sized cores, should they arise, would therefore point to the presence of a strong magnetic field. 

Finally, it is worth noting that other authors have also argued that the boundaries of convective cores are best-described by the Schwarzschild criterion on grounds that are entirely distinct from the existence of the ODDC instability. As demonstrated by \citet{Andersal22}, convective entrainment gradually pushes the location of the boundary predicted by the Ledoux criterion outwards until it agrees with the Schwarzschild criterion, on a timescale that is fast compared to stellar evolutionary timescales \citep[see also][where issues stemming from miscalculations of convective boundaries are discussed, and where the appropriateness of the Schwarzschild criterion is also argued]{Gabrielal2014, Paxtonal2018, Paxtonal2019}. Thus, there are several distinct physical arguments for using the Schwarzschild criterion over the Ledoux criterion in stellar evolution models, in the absence of magnetic fields. 
In this paper, we have proposed that there may be magnetic fields of sufficient strength to stop the weak form of ODDC from spontaneously evolving into standard convection in these regions that are Schwarzschild-unstable but Ledoux-stable.
This does not address whether the other arguments for using the Schwarzschild criterion over the Ledoux criterion still hold in MHD -- to address this question, further studies on convective entrainment in MHD are necessary.

\begin{acknowledgements}
A. S, A. F. and P. G. acknowledge funding by NSF AST 1908338. 
A. S. acknowledges funding from the Other Worlds Laboratory at UC Santa Cruz. Most simulations were run on the Lux supercomputer at UC Santa Cruz, funded by the NSF MRI grant AST-1828315. This work also used the Extreme Science and Engineering Discovery Environment (XSEDE), which is supported by National Science Foundation grant number ACI-1548562.  We thank Evan Anders and Adam Jermyn for useful discussions.
\end{acknowledgements}

\bibliographystyle{aasjournal}
\bibliography{DDClowPr_references}

\end{document}